\documentclass[
 amsmath,amssymb,
 aps,
]{revtex4-1}

\usepackage{graphicx}

\usepackage{dcolumn}

\usepackage{bm}

\begin{document}


\title{Zero Mode Effect Generalization \\ for the Electromagnetic Current in the Light Front}
\author{Alfredo Takashi Suzuki}
\affiliation{Department of Physics and Engineering, Southern Adventist University, Collegedale, TN 37315\\ 
}%
\affiliation{Instituto de F\'{\i}sica Te\'orica, UNESP-Universidade Estadual Paulista, 
Rua Dr. Bento Teobaldo Ferraz, 271, 01140-070 - S\~ao Paulo, SP, Brazil
}%

\author{Jorge Henrique Sales}
\affiliation{Universidade Estadual de Santa Cruz, Km 16 Rodovia Ilh\'eus-Itabuna, 45662-900 - Ilh\'eus, BA, Brazil
}%


\author{Luis Alberto Soriano}
\affiliation{Instituto de F\'{\i}sica Te\'orica, UNESP-Universidade Estadual Paulista,
Rua Dr. Bento Teobaldo Ferraz, 271, 01140-070 - S\~ao Paulo, SP, Brazil
}%


\date{\today}

\begin{abstract}
We consider in this work the electromagnetic current for a system composed by two charged bosons and show that it has a structure of many bodies even in the impulse approximation, when described in the light front time $x^+$. In terms of the two-body component for the bound state, the current contains two-body operators. We discuss the process of pair creation from the interacting photon and interpret it as a zero mode contribution to the current and its consequences for the components of currents in the light-front. 

\begin{description}
\item[PACS numbers] 11.10.St, 21.45.-v
\end{description}
\end{abstract}

\pacs{Valid PACS appear here}
\maketitle


\section{Introduction}

In the traditional approach to restore covariance of the electromagnetic current in the light front \cite{dirac} an ``ad hoc'' prescription of dislocating the pole is employed \cite{9}. However, this procedure of ``pole dislocation'' has no physical grounds and the arrival at the correct result is just fortuitous. We demonstrate that the light front Fock space of positive quanta solutions
is incomplete and that as a consequence the non triviality of the light front vacuum is a mandatory feature in the new scenario.

In order to demonstrate this we calculate in a specific example the matrix element for the electromagnetic current in the Breit's reference frame for $q^{+}\rightarrow 0$ and $\vec{\bf q}_{\bot}\neq 0$. To this end we use a constant vertex for the bound state of two bosons in the light front. Such calculation agrees with the results obtained through the computation of the
triangular diagram for the electromagnetic current of a composite boson whose vertex is constant \cite{9}.

Sawicki \cite{37} has shown that in the Breit's reference frame, the $J^{+}=J^{0}+J^{3}$ component of the electromagnetic current for the bound state of two bosons, obtained from the triangular Feynman diagram after integration in the $k^-$ component of the loop momentum, has no pair production contribution from the photon. As a consequence of this fact, the
electromagnetic form factor, calculated in the light front, starting from $J^+$ is identical to the one obtained in the covariant calculation. By covariant calculation of an amplitude we mean the computation of momentum loop integrated directly without the use of the transformation into light front momentum.

The problem that appears when integrating in the light front coordinates in momentum loops was studied by Chang and Yan \cite{33} and more recently discussed in references \cite{34,35,9,51}. In the Chang and Yan's works although it is pointed out the difficulty in the $k^{-}$ integration for certain amplitudes and suggested a possible solution to the problem, in our
view two distinct aspects are mingled together which are the renormalization question and the problem of integration in the light front coordinates. Our emphasis here is the covariance restoration for the electromagnetic current through a careful integration in the $k^{-}$ of the loop momentum in finite diagrams. We know that for the $J^{+}$ component of the electromagnetic current for a particle of spin $1$ there are terms that correspond to pair production in the light front formalism for $q^{+}=0$ \cite{9,51}. In the case of vector meson $\rho $, the rotational invariance of the current $J^{+} $ is broken when we use the light front formalism unless pair production diagrams are duly considered \cite{46a,41}.

In references \cite{37,38,40,39} the electromagnetic current in the light front for a composite system is obtained from the triangular diagram (impulse approximation) when this is integrated in the internal loop momentum component $k^{-}=k^{0}-k^{3}$. This integration in $k^{-}$ by Cauchy's theorem uses the pole of the spectator particle in the process of
photon absorption for $q^{+}=0$. Using the current $J^{+}$ the process of pair creation by the photon is in principle eliminated \cite{37,38}. In general, covariance is preserved under kinematic transformations, but the current looses this physical property under a more general transformation such as rotations and parity transformations. We show how the pair
production is necessary to the complete calculation of the current's $J^{-}$ component in the Drell-Yan's reference frame ($q^{+}=0$).

The paper is organized as follows: Section \ref{Prop} introduces the propagator in a background field with interacting bosons followed by a section in which we consider the relevant operator components for the electromagnetic current. Then we study the possible contributions to the order $g^2$ that might contibute to the zero modes followed by the conclusions. An appendix is added at the end of the paper that considers with some detail the case of a free propagator in the light front, which serves as a notational convention used througout this work. 
 

\section{Propagator in a Background Field with Interacting Bosons}
\label{Prop}

In a recent article \cite{Sales-Suzuki} it has been considered, in the zeroth order of perturbative coupling, the calculation of the electromagnetic current in the light front coordinates for scalar bosons in the electromagnetic background field. The calculation is considered only in the region $0<k_{2}^{+}<k_{i}^{+}<k_{4}^{+}<k_{f}^{+}$ and its combinations. The same result is found in the article by Marinho, Frederico and Sauer \cite{Marinho}, using a different technique.

The Lagrangean density for interacting scalar and electromagnetic fields is given by 
\begin{widetext}
\begin{eqnarray}
\pounds  &=&D_{\mu }\phi _{1}D^{\mu }\phi _{1}^{\ast }-m^{2}\phi _{1}^{\ast
}\phi _{1}-m^{2}\phi _{2}^{\ast }\phi _{2}+g\phi _{1}^{\ast }\phi _{1}\sigma
+g\phi _{2}^{\ast }\phi _{2}\sigma   \nonumber \\
&=&\partial _{\mu }\phi _{1}\partial ^{\mu }\phi _{1}^{\ast }-m^{2}\phi
_{1}^{\ast }\phi _{1}-m^{2}\phi _{2}^{\ast }\phi _{2}-ieA^{\mu }\left( \phi
_{1}\partial _{\mu }\phi _{1}^{\ast }-\phi _{1}^{\ast }\partial _{\mu }\phi
_{1}\right) +g\phi _{1}^{\ast }\phi _{1}\sigma +g\phi _{2}^{\ast }\phi
_{2}\sigma +e^{2}A^{\mu }A_{\mu }\phi _{1}\phi _{1}^{\ast }.
\label{L}
\end{eqnarray}
\end{widetext}

In the calculation of the propagator for a particle in a background field we use the interaction Lagrangean of a scalar field and electromagnetic field. The Lagrangean (\ref{L}) shows immediately that there are two types of vertices. The first term corresponds to a vertex containing a photon and two scalar particles. The second vertex contains two photons and two scalar
particles.

In this framework we construct the electromagnetic current operator for the system composed of two free bosons in the light front. The technique we use to deduce such operators is to define the global propagators in the light front when a electromagnetic background field acts on one of the particles. Although we are in fact calculating the global propagator for two bosons in an electromagnetic background field, we extrapolate the language using terms such as \emph{``current operator''} and \emph{``current''} to designate such operations. We show that for the $J^{-}$ case the two free boson propagators in a background field have a contribution from the process of photon's pair production, being crucial to restore current's covariance.

The normalized generating functional is given by 
\begin{equation}
Z\left[J\right] = \frac{\int \mathcal{D}\phi\, \exp \left[i\mathcal{S}+i\int dx\,J\phi \right]}{\int \mathcal{D}\phi\, e^{i\mathcal{S}}}  \label{3.3.2}
\end{equation}
where $\mathcal{S}=\int \pounds dx$ is the relevant action.  In the appendix we have the expression $Z_{0}\left[J\right]$ for the free particle and the corresponding propagator (\ref{cov}). So, we can find the propagators, or Green's functions, in an electromagnetic field.

The Eq. (\ref{3.3.2}) indicates the propagation of two bosons $S_{1}$ and $S_{2}$ that propagates from $x^{+}=0$ to $x^{+}>0$ interacting with an external electromagnetic field $A^{\mu }(x^{+})$  at $\bar{x}_{3}^{+}$ and with the exchange of two intermediate bosons $\sigma $ between $\bar{x}_{1}^{+}$ and $\bar{x}_{2}^{+}$. The propagator $S_{3}(\bar{x}_{1}^{+}-\bar{x}_{3}^{+})$ which is the propagation of a boson after the emission of boson $\sigma$ at $\bar{x}_{1}^{+}$ later interacts with the external field at $\bar{x}_{3}^{+}$. The propagator $S_{5}$ is the boson propagation after the interaction with the external field. The propagator $S_{4}$ is the boson propagation after the absorption of the intermediate $\sigma$ boson. Therefore the correction to the free propagator of two bosons in the light front with background field in the ladder diagram is: 
\begin{eqnarray}
S(x^{+}) & = & (-ie)\left(ig\right)^{2}\int d\bar{x}_{1}^{+}d\bar{x}_{2}^{+}d\bar{x}_{3}^{+}dq^{-}A^{\mu}(q^{-})\nonumber \\
         & \times & e^{-\frac{i}{2}q^{-}\bar{x}_{3}^{+}}S_{1}(\bar{x}_{1}^{+})S_{2}(\bar{x}_{2}^{+})S_{4}(x^{+}-\bar{x}_{2}^{+})  \nonumber \\
         & \times & S_{\sigma}(\bar{x}_{2}^{+}-\bar{x}_{1}^{+})\left[S_{3}\frac{\partial S_{5}}{\partial \bar{x}_{3}^{+}}\!-\!\frac{\partial S_{3}}{\partial \bar{x}_{3}^{+}}S_{5}\right] ,  \label{inicial0}
\end{eqnarray}
where $A^{\mu}(q^{-})$ is the Fourier transform and $\mu$ indicates the components $-,\, +,\, \perp$.

The part of the interaction Lagrangean that contains the vector-scalar-scalar vertex can be redefined as 
\begin{equation}
\pounds _{\text{Interaction}}=-ieA^{\mu }\left( \phi _{1}\partial_{\mu }\phi _{1}^{\ast }-\phi _{1}^{\ast }\partial _{\mu }\phi _{1}\right)=J_{\mu }A^{\mu }.  \label{campo_externo}
\end{equation}

So, we observe that the opeerator component $J^{\mu }$ is obtained from the operator $\mathcal{O}^{\mu }$ which we identified with the help of Eq. (\ref{campo_externo}) and Eq.(\ref{inicial0}), such that
\begin{eqnarray}
\mathcal{O}^{\mu } & = &(-ie)\left(ig\right)^{2} \int d\bar{x}_{1}^{+}d\bar{x}_{2}^{+}d\bar{x}_{3}^{+}e^{-\frac{i}{2}q^{-}\bar{x}_{3}^{+}} \nonumber \\
                   & \times & S_{1}(\bar{x}_{1}^{+})S_{2}(\bar{x}_{2}^{+})S_{4}(x^{+}-\bar{x}_{2}^{+})   \nonumber \\
                   & \times & S_{\sigma }(\bar{x}_{2}^{+}-\bar{x}_{1}^{+})\left[ S_{3}\frac{\partial S_{5}}{\partial \bar{x}
_{3}^{\mu }}-\frac{\partial S_{3}}{\partial \bar{x}_{3}^{\mu }}S_{5}\right] 
\text{,}  \label{op2}
\end{eqnarray}
where the Greek index $\mu$ indicates the light-front components $+,\,-,\,{\rm or} \perp$.

Therefore using the definition for the propagator and making explicit the integration and the $k^{-}$, $k^{+}$ and $k_{\perp}$ components, we have the following:
\begin{widetext}
\begin{eqnarray}
S_{1}(x^{+}) & = & \frac{i}{2(2\pi)^{4}}\int \frac{
dk_{1}^{-}dk_{1}^{+}d^{2}k_{1\bot}}{k_{1}^{+}}\frac{e^{-ik_{1}^{-}\bar{x}_{1}^{+}}e^{-ik_{1}^{+}\bar{x}_{1}^{-}}e^{i\vec{k}_{1\perp}\cdot \vec{x}_{1\perp }}}{\left[k_{1}^{-}-\frac{k_{1\bot}^{2}+m^{2}}{2k_{1}^{+}}+\frac{i\varepsilon }{2k_{1}^{+}}\right] }, 
\nonumber \\
S_{2}(x^{+}) & = &\frac{i}{2(2\pi )^{4}}\int \frac{dk_{2}^{-}dk_{2}^{+}d^{2}k_{2\bot}}{k_{2}^{+}}\frac{e^{-ik_{2}^{-}\bar{
x}_{2}^{+}}e^{-ik_{2}^{+}\bar{x}_{2}^{-}}e^{i\vec{k}_{2\perp }\cdot \vec{x}_{2\perp }}}{\left[k_{2}^{-}-\frac{k_{2\bot}^{2}+m^{2}}{2k_{2}^{+}}+\frac{i\varepsilon }{2k_{2}^{+}}\right] }, 
\nonumber \\
S_{3}(x^{+}) & = &\frac{i}{2(2\pi )^{4}}\int \frac{dk_{3}^{-}dk_{3}^{+}d^{2}k_{3\bot }}{k_{3}^{+}}\frac{e^{-ik_{3}^{-}(
\bar{x}_{3}^{+}-\bar{x}_{1}^{+})}e^{-ik_{3}^{+}(\bar{x}_{3}^{-}-\bar{x}_{1}^{-})}e^{i\vec{k}_{3\bot }\cdot(\vec{x}_{3}-\vec{x}_{1})_{\bot }}}{\left[ k_{3}^{-}-\frac{k_{3\bot }^{2}+m^{2}}{2k_{3}^{+}}+\frac{
i\varepsilon }{2k_{3}^{+}}\right] },  \nonumber \\
S_{4}(x^{+}) & = &\frac{i}{2(2\pi )^{4}}\int \frac{dk_{4}^{-}dk_{4}^{+}d^{2}k_{4\bot }}{k_{4}^{+}}\frac{e^{-ik_{4}^{-}(x^{+}-
\bar{x}_{2}^{+})}e^{-ik_{4}^{+}(x^{-}-\bar{x}_{2}^{-})}e^{i\vec{k}_{4\bot }\cdot(\vec{x}-\vec{x}_{2})_{\bot }}}{\left[ k_{4}^{-}-\frac{k_{4\bot }^{2}+m^{2}}{2k_{4}^{+}}+
\frac{i\varepsilon }{2k_{4}^{+}}\right] },  \nonumber \\
S_{5}(x^{+}) & = & \frac{i}{2(2\pi )^{4}}\int \frac{
dk_{5}^{-}dk_{5}^{+}d^{2}k_{5\bot }}{k_{5}^{+}}\frac{e^{-ik_{5}^{-}(x^{+}-
\bar{x}_{3}^{+})}e^{-ik_{5}^{+}(x^{-}-\bar{x}_{3}^{-})}e^{i
\vec{k}_{5\bot }\cdot(\vec{x}-\vec{x}_{3})_{\bot }}}{\left[ k_{5}^{-}-\frac{k_{5\bot }^{2}+m^{2}}{2k_{5}^{+}}+
\frac{i\varepsilon }{2k_{5}^{+}}\right] },  \nonumber \\
S_{\sigma }(x^{+}) & = & \frac{i}{2(2\pi )^{4}}\int \frac{dk_{\sigma
}^{-}dk_{\sigma }^{+}d^{2}k_{\sigma \bot }}{k_{\sigma }^{+}}\frac{e^{-ik_{\sigma }^{-}(\bar{x}_{2}^{+}-\bar{x}_{1}^{+})}e^{-ik_{
\sigma }^{+}(\bar{x}_{2}^{-}-\bar{x}_{1}^{-})}e^{i\vec{k}_{\sigma \bot }\cdot (\vec{x}_{2}-\vec{x}_{1})_{\bot }}
}{\left[ k_{\sigma }^{-}-\frac{k_{\sigma \bot }^{2}+m^{2}}{2k_{\sigma }^{+}}+
\frac{i\varepsilon }{2k_{\sigma }^{+}}\right] },  \label{momentos}
\end{eqnarray}
\end{widetext}

For the bosons identified by the labels $3$ and $5$ we also need the derivatives with respect to $\bar{x}_3^{\mu}$ as follows:
\begin{itemize}
\item With respect to component $\bar{x}_{3}^{+}:$
\end{itemize}
\begin{widetext}
\begin{eqnarray}
\frac{\partial S_{5}}{\partial \bar{x}_{3}^{+}} & = &-\frac{1}{2(2\pi )^{4}}\int \frac{dk_{5}^{-}dk_{5}^{+}d^{2}k_{5\bot }}{k_{5}^{+}}\frac{k_{5}^{-}e^{-ik_{5}^{-}(x^{+}-\bar{x}_{3}^{+})}e^{-ik_{5}^{+}(x^{-}-\bar{x}_{3}^{-})}e^{i\vec{k}_{5\bot }\cdot(\vec{x}-\vec{\bar{x}}_{3})_{\bot }}}{\left[ k_{5}^{-}-\frac{k_{5\bot
}^{2}+m^{2}}{2k_{5}^{+}}+\frac{i\varepsilon }{2k_{5}^{+}}\right] }, 
\nonumber \\
\frac{\partial S_{3}}{\partial \bar{x}_{3}^{+}} &=&\frac{1}{2(2\pi )^{4}}\int \frac{dk_{3}^{-}dk_{3}^{+}d^{2}k_{3\bot }}{k_{3}^{+}}\frac{k_{3}^{-}e^{-ik_{3}^{-}(\bar{x}_{3}^{+}-\bar{x}_{1}^{+})}e^{-ik_{3}^{+}(\bar{x}_{3}^{-}-\bar{x}_{1}^{-})}e^{i\vec{k}_{3\bot }\cdot(\vec{\bar{x}}_{3}-\vec{\bar{x}}_{1})_{\bot }}}{\left[ k_{3}^{-}-\frac{k_{3\bot
}^{2}+m^{2}}{2k_{3}^{+}}+\frac{i\varepsilon }{2k_{3}^{+}}\right] }.
\end{eqnarray}
\end{widetext}

\begin{itemize}
\item With respect to component $\bar{x}_{3}^{-}:$
\end{itemize}
\begin{widetext}
\begin{eqnarray}
\frac{\partial S_{5}}{\partial \bar{x}_{3}^{-}} & = &-\frac{1}{2(2\pi )^{4}}\int \frac{dk_{5}^{-}dk_{5}^{+}d^{2}k_{5\bot }}{k_{5}^{+}}\frac{k_{5}^{+}e^{-ik_{5}^{-}(x^{+}-\bar{x}_{3}^{+})}e^{-ik_{5}^{+}(x^{-}-\bar{x}_{3}^{-})}e^{i\vec{k}_{5\bot }\cdot(\vec{x}-\vec{\bar{x}}_{3})_{\bot }}}{\left[ k_{5}^{-}-\frac{k_{5\bot}^{2}+m^{2}}{2k_{5}^{+}}+\frac{i\varepsilon }{2k_{5}^{+}}\right] }, 
\nonumber \\
\frac{\partial S_{3}}{\partial \bar{x}_{3}^{-}} &=&\frac{1}{2(2\pi )^{4}}\int \frac{dk_{3}^{-}dk_{3}^{+}d^{2}k_{3\bot}}{k_{3}^{+}}\frac{k_{3}^{+}e^{-ik_{3}^{-}(\bar{x}_{3}^{+}-\bar{x}_{1}^{+})}e^{-ik_{3}^{+}(\bar{x}_{3}^{-}-\bar{x}_{1}^{-})}e^{i\vec{k}_{3\bot }\cdot(\vec{\bar{x}}_{3}-\vec{\bar{x}}_{1})_{\bot }}}{\left[ k_{3}^{-}-\frac{k_{3\bot}^{2}+m^{2}}{2k_{3}^{+}}+\frac{i\varepsilon }{2k_{3}^{+}}\right] }.
\end{eqnarray}
\end{widetext}

\begin{itemize}
\item With respect to component $\bar{x}_{3\perp }:$
\end{itemize}
\begin{widetext}
\begin{eqnarray}
\frac{\partial S_{5}}{\partial \bar{x}_{3\perp }} & = &\frac{1}{2(2\pi )^{4}}\int \frac{dk_{5}^{-}dk_{5}^{+}d^{2}k_{5\perp }}{k_{5}^{+}}\frac{k_{5\perp}e^{-ik_{5}^{-}(x^{+}-\bar{x}_{3}^{+})}e^{-ik_{5}^{+}(x^{-}-\bar{x}_{3}^{-})}e^{i\vec{k}_{5\perp }\cdot(\vec{x}-\vec{\bar{x}}_{3})_{\perp }}}{\left[ k_{5}^{-}-\frac{k_{5\perp }^{2}+m^{2}}{2k_{5}^{+}}+\frac{i\varepsilon }{2k_{5}^{+}}\right] },  \nonumber \\
\frac{\partial S_{3}}{\partial \bar{x}_{3\perp }} & = &-\frac{1}{2(2\pi )^{4}}\int \frac{dk_{3}^{-}dk_{3}^{+}d^{2}k_{3\perp }}{k_{3}^{+}}\frac{k_{3\perp}e^{-ik_{3}^{-}(\bar{x}_{3}^{+}-\bar{x}_{1}^{+})}e^{-ik_{3}^{+}(\bar{x}_{3}^{-}-\bar{x}_{1}^{-})}e^{i\vec{k}_{3\perp }\cdot(\overrightarrow{\bar{x}}_{3}-\overrightarrow{\bar{x}}_{1})_{\perp }}}{\left[ k_{3}^{-}-\frac{k_{3\perp }^{2}+m^{2}}{2k_{3}^{+}}+\frac{i\varepsilon }{2k_{3}^{+}}\right] }.
\end{eqnarray}
\end{widetext}
Substituting Eq. (\ref{momentos}) and the relevant derivatives above in Eq. (\ref{inicial0}) and performing the integrations over $d\bar{x}_{1}^{+}d\bar{x}_{2}^{+}d\bar{x}_{3}^{+}$ we can evaluate the Fourier transform $\widetilde{S}(k_{f}^{-}) = \int dx^{+}e^{ik_{f}^{-}x^{+}}S(x^{+})$ with the help of the following momentum conservation relations:
\begin{eqnarray}
k_i = k_f \quad & ; & \quad k_{i} = k_{1}+k_{2} \nonumber  \\
k_{f} = k_{4}+k_{5} \quad & ; & \quad  k_{f} = k_{i}+q \nonumber \\
k_{3} = k_{i}-k_{4} \quad  & ; & \quad q = k_{5}-k_{3} \nonumber \\
k_{\sigma} = k_{4}-k_{2} \quad & ; & \quad k_{\sigma } = k_{1}-k_{3}. 
\label{mom_cons}
\end{eqnarray}
where $k_i$ is the total initial momentum and $k_f$, the total final momentum.

The final propagator can therefore be written as a function of only two momenta, and in this case we choose ``spectator'' particles with respect to the current, those labeled as $2$ and $4$:  
\begin{widetext}
\begin{eqnarray}
\widetilde{S}(k_{f}^{-}) & = &-\frac{ie\left( ig\right)^{2}}{2^{6}(2\pi )^{2}}\int dq^{-}\,A^{\mu}(q^-)\left\{ \int \frac{dk_{2}^{-}dk_{4}^{-}\left( k_{f}^{\mu}+k_{i}^{\mu}-2k_{4}^{\mu }\right) }{\left( k_{i}-k_{2}\right)^{+}k_{2}^{+}\left( k_{i}-k_{4}\right) ^{+}k_{4}^{+}\left(k_{f}-k_{4}\right)^{+}\left( k_{4}-k_{2}\right)^{+}} \right.  
\nonumber \\
& \times & \frac{1}{\left[ k_{2}^{-}-k_{i}^{-}+(k_{i}-k_{2})_{\text{on}}-\frac{i\epsilon }{2\left( k_{i}-k_{2}\right) ^{+}}\right] \left[k_{2}^{-}-k_{2\text{on}}+\frac{i\epsilon }{2k_{2}^{+}}\right] }   \nonumber \\
& \times & \frac{1}{\left[ k_{4}^{-}-k_{i}^{-}+(k_{i}-k_{4})_{\text{on}}-\frac{i\epsilon }{2\left(k_{i}-k_{4}\right) ^{+}}\right] \left[k_{4}^{-}-k_{4\text{on}}+\frac{i\epsilon }{2k_{4}^{+}}\right] }  \nonumber \\
& \times & \left.\frac{1}{\left[k_{2}^{-}-k_{4}^{-}+(k_{4}-k_{2})_{\text{on}}-\frac{i\epsilon }{2\left(k_{4}-k_{2}\right) ^{+}}\right]\left[ k_{4}^{-}-k_{f}^{-}+(k_{f}-k_{4})_{\text{on}}-\frac{i\epsilon }{2\left(k_{f}-k_{4}\right)^{+}}\right] } \right\} \text{,}  \label{intermediario}  
\end{eqnarray}
\end{widetext}
where
\begin{widetext}
\begin{eqnarray}
k_{2\text{on}} = \frac{k_{2_{\perp }}^{2}+m^{2}}{2k_{2}^{+}} \quad & ; & \quad (k_{i}-k_{2})_{\text{on}} = \frac{(k_{i}-k_{2})_{\perp }^{2}+m^{2}}{2(k_{i}^{+}-k_{2}^{+})} \nonumber \\
(k_{i}-k_{4})_{\text{on}} = \frac{(k_{i}-k_{4})_{\perp }^{2}+m^{2}}{2(k_{i}^{+}-k_{4}^{+})} \quad & ; & \quad k_{4\text{on}} = \frac{k_{4_{\perp }}^{2}+m^{2}}{2k_{4}^{+}} \nonumber \\ 
(k_{f}-k_{4})_{\text{on}} = \frac{(k_{f}-k_{4})_{\perp }^{2}+m^{2}}{2(k_{i}^{+}-k_{4}^{+})}\quad & ; & \quad (k_{4}-k_{2})_{\text{on}} = \frac{(k_{4}-k_{2})_{\perp }^{2}+m_{\sigma }^{2}}{2(k_{4}^{+}-k_{2}^{+})}\text{.}
\end{eqnarray}
\end{widetext}

We begin our discussion with an illustrative example, where the pair term appears. To this end we use the ``Z-graph'', that is, the region 3 of momentum possibilities $0 < k_{2}^{+}<k_{i}^{+}<k_{4}^{+}<k_{f}^{+}$. In this example we show that the current's $J^{-}$ component does not have a contribution from the pair production in the Drell-Yan reference frame \cite{Drell}, that is, in the limit $q^{+}=q^{-}=0$.

We look after the components of the current operator $J^{\mu }$, which as we referred before, will be obtained from the operator $\mathcal{O}^{\mu }$ which is represented by the square brackets in Eq. (\ref{inicial0}), so
\begin{widetext}
\begin{eqnarray}
\mathcal{O}_{3}^{\mu } &=&-\frac{ie\left( ig\right) ^{2}}{2^{6}(2\pi )^{2}}\int \frac{dk_{2}^{-}dk_{4}^{-}\left( k_{f}^{\mu }+k_{i}^{\mu }-2k_{4}^{\mu}\right) }{\left( k_{i}-k_{2}\right) ^{+}k_{2}^{+}\left( k_{i}-k_{4}\right)
^{+}k_{4}^{+}\left( k_{f}-k_{4}\right) ^{+}\left( k_{4}-k_{2}\right) ^{+}}   \nonumber \\
&&\times \frac{1}{\left[ k_{2}^{-}-k_{i}^{-}+(k_{i}-k_{2})_{\rm on}-\frac{i\epsilon }{2\left( k_{i}-k_{2}\right) ^{+}}\right] \left[k_{2}^{-}-k_{2\rm on}+\frac{i\epsilon }{2k_{2}^{+}}\right] }   \nonumber\\
&&\times \frac{1}{\left[ k_{4}^{-}-k_{i}^{-}+(k_{i}-k_{4})_{\rm on}-\frac{i\epsilon }{2\left( k_{i}-k_{4}\right) ^{+}}\right] \left[k_{4}^{-}-k_{4\rm on}+\frac{i\epsilon }{2k_{4}^{+}}\right] }   \nonumber\\
&&\times \frac{1}{\left[ k_{2}^{-}-k_{4}^{-}+(k_{4}-k_{2})_{\rm on}-\frac{i\epsilon }{2\left( k_{4}-k_{2}\right) ^{+}}\right] \left[ k_{4}^{-}-k_{f}^{-}+(k_{f}-k_{4})_{\rm on}-\frac{i\epsilon }{2\left( k_{f}-k_{4}\right) ^{+}}\right]} \text{.}
\label{Opcorrente1}
\end{eqnarray}
\end{widetext}
where $k_{i}^{\mu }$ and $k_{f}^{\mu }$ are the initial and final four-momentum of the system and $m$ is the mass of the boson. The integration in Eq. (\ref{Opcorrente1}), using the Cauchy integral formula over $k_{2}^{-}$ and $k_{4}^{-}$, have ten nonvanishing contributions for the residue calculation, but for our example, we concentrate our attention in a specific region, that is, in the range of momenta satisfying $0<k_{2}^{+}<k_{i}^{+}<k_{4}^{+}<k_{f}^{+}$, which corresponds to the ``Z-graph'' or what we call region 3. Thus we are ready for working out the distinct components of the operator, which we detail in the next section.

\section{Operator Components $\mathcal{O}^{-,+,\perp}$}

After performing the relevant integrals via Cauchy's residue theorem in Eq. (\ref{Opcorrente1}), we have for the $-$ component: 
\begin{widetext}
\begin{eqnarray}
\mathcal{O}_{3}^{-} & = & \frac{ie\left(ig\right)^{2}}{2^{6}}\frac{\theta\left(k_{i}^{+}-k_{2}^{+}\right) \theta \left(k_{2}^{+}\right) \theta\left(k_{4}^{+}-k_{i}^{+}\right) \theta \left(k_{4}^{+}\right) \theta\left(k_{f}^{+}-k_{4}^{+}\right) \theta \left(k_{4}^{+}-k_{2}^{+}\right)}{\left(k_{i}-k_{2}\right)^{+}k_{2}^{+}\left( k_{i}-k_{4}\right)^{+}k_{4}^{+}\left(k_{f}-k_{4}\right)^{+}\left(k_{4}-k_{2}\right)^{+}}  \nonumber \\
                    & \times & \frac{\left[k_{f}^{-}-k_{i}^{-}-\frac{(k_{f}-k_{4})_{\bot}^{2}+m^{2}}{(k_{f}-k_{4})}\right]}{\left[ k_{i}^{-}-\frac{
(k_{i}-k_{2})_{\bot }^{2}+m^{2}}{2(k_{i}^{+}-k_{2}^{+})}-\frac{k_{2}^{2}{}_{\bot }+m^{2}}{2k_{2}^{+}}\right] }\frac{1}{\left[ k_{f}^{-}-\frac{k_{2}^{2}{}_{\bot }+m^{2}}{2k_{2}^{+}}-\frac{(k_{f}-k_{4})_{\bot}^{2}+m^{2}}{2(k_{f}^{+}-k_{4}^{+})}-\frac{(k_{4}-k_{2})_{\bot }^{2}+m_{\sigma }^{2}}{2(k_{4}^{+}-k_{2}^{+})}\right]}   \nonumber \\
                    & \times & \frac{1}{\left[ k_{f}^{-}-k_{i}^{-}+\frac{(k_{i}-k_{4})_{\bot}^{2}+m^{2}}{2(k_{i}^{+}-k_{4}^{+})}-\frac{(k_{f}-k_{4})_{\bot }^{2}+m^{2}}{2(k_{f}^{+}-k_{4}^{+})}\right]\left[ k_{f}^{-}-\frac{k_{4}^{2}{}_{\bot }+m^{2}}{2k_{4}^{+}}-\frac{(k_{f}-k_{4})_{\bot }^{2}+m^{2}}{2(k_{f}^{+}-k_{4}^{+})}\right] }\text{.}   \label{denscorrent1}
\end{eqnarray}
\end{widetext}

The physical process represented by Eq. (\ref{denscorrent1}) is the pair creation due to the interacting photon. The denominator
$$\left[k_{f}^{-}-k_{i}^{-}+\frac{(k_{i}-k_{4})_{\bot }^{2}+m^{2}}{2(k_{i}^{+}-k_{4}^{+})}-\frac{(k_{f}-k_{4})_{\bot }^{2}+m^{2}}{2(k_{f}^{+}-k_{4}^{+})}\right] ^{-1}$$
corresponds to the propagation in the intermediate state of a pair particle-antiparticle, composed by the initial bound state, the particle and the antiparticle produced by the photon. The denominators 
$$\left[ k_{i}^{-}-\frac{(k_{i}-k_{2})_{\bot }^{2}+m^{2}}{2(k_{i}^{+}-k_{2}^{+})}-\frac{k_{2}^{2}{}_{\bot}+m^{2}}{2k_{2}^{+}}\right] ^{-1}\,,$$ 
$$\left[ k_{f}^{-}-\frac{k_{2}^{2}{}_{\bot }+m^{2}}{2k_{2}^{+}}-\frac{(k_{f}-k_{4})_{\bot}^{2}+m^{2}}{2(k_{f}^{+}-k_{4}^{+})}-\frac{(k_{4}-k_{2})_{\bot}^{2}+m_{\sigma }^{2}}{2(k_{4}^{+}-k_{2}^{+})}\right]^{-1}$$
and 
$$\left[k_{f}^{-}-\frac{k_{4}^{2}{}_{\bot }+m^{2}}{2k_{4}^{+}}-\frac{(k_{f}-k_{4})_{\bot }^{2}+m^{2}}{2(k_{f}^{+}-k_{4}^{+})}\right]^{-1}$$
are the intermediate states of two and three particles that propagates forward in time $x^+$.

In a similar way, we obtain for the components $\mathcal{O}_{3}^{+}$ and  $\mathcal{O}_{3}^{\bot }$ which are given by:
\begin{widetext}
\begin{eqnarray}
\mathcal{O}_{3}^{\beta } &=&-\frac{ie\left( ig\right) ^{2}}{2^{6}}\frac{\theta \left( k_{i}^{+}-k_{2}^{+}\right) \theta \left( k_{2}^{+}\right)\theta \left( k_{4}^{+}-k_{i}^{+}\right) \theta \left( k_{4}^{+}\right)\theta \left(k_{f}^{+}-k_{4}^{+}\right) \theta \left(k_{4}^{+}-k_{2}^{+}\right) }{\left( k_{i}^{+}-k_{2}^{+}\right)k_{2}^{+}\left( k_{i}^{+}-k_{4}^{+}\right)k_{4}^{+}\left(k_{f}^{+}-k_{4}^{+}\right) \left( k_{4}^{+}-k_{2}^{+}\right) } \nonumber \\
&& \times \frac{\left[k_{f}^{\beta }+k_{i}^{\beta }-2k_{4}^{\beta }\right] }{\left[ k_{i}^{-}-\frac{(k_{i}-k_{2})_{\bot }^{2}+m^{2}}{2(k_{i}^{+}-k_{2}^{+})}-\frac{k_{2}^{2}{}_{\bot }+m^{2}}{2k_{2}^{+}}\right] } \frac{1}{\left[ k_{f}^{-}-\frac{k_{2}^{2}{}_{\bot}+m^{2}}{2k_{2}^{+}}-\frac{(k_{f}-k_{4})_{\bot}^{2}+m^{2}}{2(k_{f}^{+}-k_{4}^{+})}-\frac{(k_{4}-k_{2})_{\bot }^{2}+m_{\sigma }^{2}}{2(k_{4}^{+}-k_{2}^{+})}\right] }  \nonumber \\
&& \times \frac{1}{\left[k_{f}^{-}-k_{i}^{-}+\frac{(k_{i}-k_{4})_{\bot}^{2}+m^{2}}{2(k_{i}^{+}-k_{4}^{+})}-\frac{(k_{f}-k_{4})_{\bot }^{2}+m^{2}}{2(k_{f}^{+}-k_{4}^{+})}\right]\left[ k_{f}^{-}-\frac{k_{4}^{2}{}_{\bot }+m^{2}}{2k_{4}^{+}}-\frac{(k_{f}-k_{4})_{\bot}^{2}+m^{2}}{2(k_{f}^{+}-k_{4}^{+})}\right] }  \text{,}  \label{denscorrent2}
\end{eqnarray}
\end{widetext}
where we have introduced the notation $\beta = +,\, \perp$. The difference between the operators $\mathcal{O}_{3}^{-}$ and $\mathcal{O}_{3}^{\beta }$ is in the numerator of Eq. (\ref{denscorrent1}) and Eq. (\ref{denscorrent2}), which has components $+$ and $\bot $ instead of $-$.

\section{Zero Mode Contribution at $\mathcal{O}(g^{2})$}

To calculate the electromagnetic current generated by the diverse configurations we must have the matrix elements $J^{-,+,\perp }=\left\langle\Gamma \right. |\mathcal{O}^{-,+,\perp }|\left. \Gamma \right\rangle $, where $\Gamma$ is the constant vertex and $\mathcal{O}^{-,+,\perp}$ are the current operator components, which we can obtain directly from the sum of the final results in each region. Introducing the unit resolution into the matrix element we have:
\begin{widetext}
\begin{eqnarray}
\left\langle \Gamma \left\vert \mathcal{O}^{-,+,\bot }\right\vert \Gamma\right\rangle  & = & \int dk_{j}^{+}d^{2}k_{j\perp} \left\langle \Gamma \left\vert k_{j}^{+},\,\vec{k}_{j\perp} \right\rangle
\left\langle k_{j}^{+},\vec{k}_{j\perp} \right\vert \mathcal{O}^{-,+,\bot }\int dk_{j}^{{\prime }+}d^{2}k_{j\bot }^{\prime}\left\vert k_{j}^{{\prime }+},\,\vec{k}_{j\bot }^{\prime } \right\rangle \left\langle k_{j}^{{\prime }+},\,\vec{k}_{j\bot}^{\prime } \right\vert \Gamma \right\rangle \nonumber \\
& = & \Gamma \int dk_{j}^{+}d^{2}k_{j_{\perp }}dk_{j}^{{\prime}+}d^{2}k_{j\bot }^{\prime }\left\langle k_{j}^{+},\vec{k}_{j\perp}\right\vert \mathcal{O}^{-,+,\bot }\left\vert k_{j}^{{\prime }+},\vec{k}_{j\bot }^{\prime}\right\rangle \Gamma   \nonumber \\
&=&\Gamma^{2}\int dk_{j}^{+}d^{2}k_{j{\perp }}dk_{j}^{{\prime}+}d^{2}k_{j\bot }^{\prime }\delta (k_{j}^{+}\!-\!k_{j}^{{\prime
}+}\!-\!q^{+})\delta (\vec{k}_{j{\perp }}\!-\!\vec{k}_{j\bot}^{\prime }\!-\!\vec{q}_{\bot })\left\langle k_{j}^{+},\vec{k}_{j\perp}\right\vert 
\mathcal{O}^{-,+,\bot }\left\vert k_{j}^{{\prime }+},\vec{k}_{j\bot }^{\prime }\right\rangle   \nonumber \\
&=&\Gamma^{2}\int dk_{2}^{+}d^{2}k_{2{\perp }}dk_{4}^{+}d^{2}k_{4{\perp }}\mathcal{O}^{-,+,\bot }\text{,} \label{matrixel}
\end{eqnarray}
\end{widetext}

Thus, for the example of Fig.(4.1) the electromagnetic current $J_{3}^{-,+,\perp }$, pertinent to the region 3 with momenta range $0<k_{2}^{+}<k_{i}^{+}<k_{4}^{+}<k_{f}^{+}$ is obtained by substituting $\mathcal{O}_{3}^{-}$ given in Eq. (\ref{denscorrent1}), and $\mathcal{O}_{3}^{\beta}$ given in Eq. (\ref{denscorrent2}),

Our next step is to perform the remaining momentum integration over $k_{2}^{+}$ and $k_{4}^{+}$ and take the limit $q^{+} \rightarrow 0$. To calculate the momentum integrations we make two changes of variables that will facilitate our job of integrating them 
\begin{eqnarray}
x &=&\frac{k_{i}^{+}-k_{2}^{+}}{q^{+}}  \nonumber \\
y &=&\frac{k_{f}^{+}-k_{4}^{+}}{q^{+}}\text{.}  \label{mud var}
\end{eqnarray}

On the other hand, taking advantage of the momentum conservation relations in Eq. (\ref{mom_cons}) we get
\begin{eqnarray}
k_{1}^{+} &=&xq^{+}  \nonumber \\
k_{3}^{+} &=&\left( y-1\right) q^{+}  \nonumber \\
k_{5}^{+} &=&yq^{+}  \nonumber \\
k_{\sigma }^{+} &=&\left( x-y+1\right) q^{+}\text{.}  \label{mud var 2}
\end{eqnarray}

Now it is just a matter of putting things together. We begin by taking the $-$ component:
\begin{itemize}
\item \textbf{Current} $J_{3}^{-}$
\end{itemize}

Substituting Eqs. (\ref{mud var}) and (\ref{mud var 2}) in Eq. (\ref{matrixel} we get:
\begin{eqnarray}
J_{3}^{-} &=&\left\langle \Gamma \left\vert \mathcal{O}_{3}^{-}\right\vert
\Gamma \right\rangle =\Gamma ^{2}\int dk_{2}^{+}d^{2}k_{2_{\perp
}}dk_{4}^{+}d^{2}k_{4_{\perp }}\mathcal{O}^{-}  \nonumber \\
&=&\Gamma ^{2}\int d^{2}k_{2_{\perp }}d^{2}k_{4_{\perp }}\left\{ \left(
q^{+}\right) ^{2}\int dxdy\mathcal{O}_{3}^{-}\right\} \text{,}  \label{shif}
\end{eqnarray}
where the operator contribution $\mathcal{O}_{3}^{-}$ takes the following form
\begin{widetext}
\begin{eqnarray}
\mathcal{O}_{3}^{-} & = & \frac{ie\left( ig\right)^{2}}{2^{6}}\frac{\theta\left( k_{i}^{+}-k_{2}^{+}\right) \theta \left( k_{2}^{+}\right) \theta\left( k_{4}^{+}-k_{i}^{+}\right) \theta \left( k_{4}^{+}\right) \theta \left( k_{f}^{+}-k_{4}^{+}\right) \theta \left( k_{4}^{+}-k_{2}^{+}\right) }{xq^{+}\left( k_{i}^{+}-xq^{+}\right) \left( y-1\right) q^{+}\left(k_{f}^{+}-yq^{+}\right) yq^{+}\left( x-y+1\right) q^{+}}   \nonumber \\
&&\times \frac{\left[ k_{f}^{-}-k_{i}^{-}-\frac{(k_{f}-k_{4})_{\bot}^{2}+m^{2}}{yq^{+}}\right] }{\left[k_{i}^{-}-\frac{(k_{i}-k_{2})_{\bot}^{2}+m^{2}}{2xq^{+}}-\frac{k_{2_{\bot }}^{2}+m^{2}}{2\left(k_{i}^{+}-xq^{+}\right) }\right] }\frac{1}{\left[ k_{f}^{-}-\frac{k_{2_{\bot }}^{2}+m^{2}}{2\left(k_{i}^{+}-xq^{+}\right) }-\frac{(k_{f}-k_{4})_{\bot }^{2}+m^{2}}{2yq^{+}}-\frac{(k_{4}-k_{2})_{\bot }^{2}+m_{\sigma }^{2}}{2\left( x-y+1\right) q^{+}}\right] }\times  \nonumber \\
&&\times \frac{1}{\left[ k_{f}^{-}-k_{i}^{-}+\frac{(k_{i}-k_{4})_{\bot}^{2}+m^{2}}{2\left( y-1\right)q^{+}}-\frac{(k_{f}-k_{4})_{\bot }^{2}+m^{2}}{2yq^{+}}\right] } \frac{1}{\left[ k_{f}^{-}-\frac{k_{4_{\bot }}^{2}+m^{2}}{2\left(k_{f}^{+}-yq^{+}\right) }-\frac{(k_{f}-k_{4})_{\bot }^{2}+m^{2}}{2yq^{+}}\right] }\text{,}  \label{primeira forma}
\end{eqnarray}
\end{widetext}
which can be written in a more convenient form, factoring out all the relevant factors of $q^{+}$ to make more evident how this particular operator component depends on $q^{+}$
\begin{widetext}
\begin{eqnarray}
{\mathcal{O}}_{3}^{-} & = &\frac{ie\left(ig\right)^{2}}{2^{6}}\left(\frac{1}{q^{+}}\right) \frac{\theta \left(k_{i}^{+}-k_{2}^{+}\right) \theta\left( k_{2}^{+}\right) \theta \left( k_{4}^{+}-k_{i}^{+}\right) \theta
\left( k_{4}^{+}\right) \theta \left( k_{f}^{+}-k_{4}^{+}\right) \theta\left( k_{4}^{+}-k_{2}^{+}\right) }{x\left( k_{i}^{+}-xq^{+}\right) \left(y-1\right) \left( k_{f}^{+}-yq^{+}\right) y\left( x-y+1\right) } \nonumber \\
&&\times \frac{\left[ \left( k_{f}^{-}-k_{i}^{-}\right) q^{+}-\frac{(k_{f}-k_{4})_{\bot }^{2}+m^{2}}{y}\right] }{\left[ k_{i}^{-}q^{+}-\frac{(k_{i}-k_{2})_{\bot }^{2}+m^{2}}{2x}-\frac{k_{2_{\bot }}^{2}+m^{2}}{2\left(k_{i}^{+}-xq^{+}\right) }q^{+}\right] }\frac{1}{\left[ k_{f}^{-}q^{+}-\frac{k_{2_{\bot }}^{2}+m^{2}}{2\left(k_{i}^{+}-xq^{+}\right)}q^{+}-\frac{(k_{f}-k_{4})_{\bot }^{2}+m^{2}}{2y}-\frac{(k_{4}-k_{2})_{\bot }^{2}+m_{\sigma }^{2}}{2\left( x-y+1\right) }\right] }  \nonumber \\
&&\times \frac{1}{\left[ \left( k_{f}^{-}-k_{i}^{-}\right) q^{+}+\frac{(k_{i}-k_{4})_{\bot }^{2}+m^{2}}{2\left( y-1\right) }-\frac{(k_{f}-k_{4})_{\bot }^{2}+m^{2}}{2y}\right]\left[ k_{f}^{-}q^{+}-\frac{k_{4_{\bot }}^{2}+m^{2}}{%
2\left( k_{f}^{+}-yq^{+}\right) }q^{+}-\frac{(k_{f}-k_{4})_{\bot }^{2}+m^{2}%
}{2y}\right] }\text{.}  \label{forma conveniente}
\end{eqnarray}
\end{widetext}

We observe that substitution of Eq. (\ref{forma conveniente}) into Eq. (\ref{shif}), implies that the matrix element for $J_{3}^{-}$ is now proportional to $q^{+}$. Then, in the Drell-Yan reference frame \cite{Drell}, $q^{+}=q^{-}\rightarrow 0$, the current component $J_{3}^{-}$ does vanish. 

\begin{itemize}
\item \textbf{Current} $J_{3}^{+}$
\end{itemize}

In a similar way the component $J_{3}^{+}$ can be worked out and yields
\begin{eqnarray}
J_{3}^{+} &=&\left\langle \Gamma \left\vert \mathcal{O}_{3}^{+}\right\vert\Gamma \right\rangle =\Gamma ^{2}\int dk_{2}^{+}d^{2}k_{2_{\perp}}dk_{4}^{+}d^{2}k_{4_{\perp }}\mathcal{O}^{+}  \nonumber \\
&=&\Gamma ^{2}\int d^{2}k_{2_{\perp }}d^{2}k_{4_{\perp }}\left\{ \left(q^{+}\right) ^{2}\int dxdy\mathcal{O}_{3}^{+}\right\} \text{,}  \label{shif1}
\end{eqnarray}
where the operator factor ${\mathcal{O}}_{3}^{+}$ is now
\begin{widetext}
\begin{eqnarray}
\mathcal{O}_{3}^{+} & = &\frac{i\left( ig\right) ^{2}}{2^{6}}\frac{\theta\left( k_{i}^{+}-k_{2}^{+}\right) \theta \left( k_{2}^{+}\right) \theta\left( k_{4}^{+}-k_{i}^{+}\right) \theta \left( k_{4}^{+}\right) \theta\left( k_{f}^{+}-k_{4}^{+}\right) \theta \left( k_{4}^{+}-k_{2}^{+}\right) }{x\left( k_{i}^{+}-xq^{+}\right) \left( y-1\right) \left(
k_{f}^{+}-yq^{+}\right) y\left( x-y+1\right) }  \nonumber \\
&&\times \frac{q^{+}\left( 2y-1\right) }{\left[ k_{i}^{-}q^{+}-\frac{(k_{i}-k_{2})_{\bot }^{2}+m^{2}}{2x}-\frac{k_{2_{\bot }}^{2}+m^{2}}{2\left(k_{i}^{+}-xq^{+}\right) }q^{+}\right] }\frac{1}{\left[ k_{f}^{-}q^{+}-\frac{k_{2_{\bot }}^{2}+m^{2}}{2\left( k_{i}^{+}-xq^{+}\right) }q^{+}-\frac{(k_{f}-k_{4})_{\bot }^{2}+m^{2}}{2y}-\frac{(k_{4}-k_{2})_{\bot }^{2}+m_{\sigma }^{2}}{2\left( x-y+1\right) }\right] }  \nonumber \\
&&\times \frac{1}{\left[ \left( k_{f}^{-}-k_{i}^{-}\right) q^{+}+\frac{(k_{i}-k_{4})_{\bot }^{2}+m^{2}}{2\left( y-1\right) }-\frac{(k_{f}-k_{4})_{\bot }^{2}+m^{2}}{2y}\right]\left[ k_{f}^{-}q^{+}-\frac{k_{4_{\bot }}^{2}+m^{2}}{2\left( k_{f}^{+}-yq^{+}\right) }q^{+}-\frac{(k_{f}-k_{4})_{\bot }^{2}+m^{2}}{2y}\right] } \text{,}  \label{Conveniente 2}
\end{eqnarray}
\end{widetext}
where we have used Eqs. (\ref{mud var}) and (\ref{mud var 2}) and found that $\left[k_{f}^{+}+k_{i}^{+}-2k_{4}^{+}\right] =q^{+}\left(2y-1\right)$. We observe that in the substitution of Eq. (\ref{Conveniente 2}) into Eq. (\ref{shif1}) the current component $J_{3}^{+}$ is proportional to $\left(q^{+}\right)^{3}$. Then, in the Drell-Yan frame of reference, the current $J^{+}$ also vanishes. 

\begin{itemize}
\item \textbf{Current }$J_{3}^{\bot }$ 
\end{itemize}

In the same manner, for $J_{3}^{\bot }$ we have
\begin{eqnarray}
J_{3}^{\bot } & = &\left\langle \Gamma \left\vert \mathcal{O}_{3}^{\bot
}\right\vert \Gamma \right\rangle =\Gamma ^{2}\int
dk_{2}^{+}d^{2}k_{2_{\perp }}dk_{4}^{+}d^{2}k_{4_{\perp }}\mathcal{O}^{\bot }
\nonumber \\
&=&\Gamma ^{2}\int d^{2}k_{2_{\perp }}d^{2}k_{4_{\perp }}\left\{ \left(
q^{+}\right) ^{2}\int dxdy\mathcal{O}_{3}^{\bot }\right\} \text{,}
\label{shif2}
\end{eqnarray}
where the operator factor $\mathcal{O}_{3}^{\bot }$ has the following form
\begin{widetext}
\begin{eqnarray}
\mathcal{O}_{3}^{\bot } &=&\frac{i\left( ig\right) ^{2}}{2^{6}}\frac{\theta\left( k_{i}^{+}-k_{2}^{+}\right) \theta \left( k_{2}^{+}\right) \theta\left( k_{4}^{+}-k_{i}^{+}\right) \theta \left( k_{4}^{+}\right) \theta\left( k_{f}^{+}-k_{4}^{+}\right) \theta \left( k_{4}^{+}-k_{2}^{+}\right) }{x\left( k_{i}^{+}-xq^{+}\right) \left( y-1\right) \left(
k_{f}^{+}-yq^{+}\right) y\left( x-y+1\right) }  \nonumber \\
&&\times \frac{\left[ k_{f}^{\bot }+k_{i}^{\bot }-2k_{4}^{\bot }\right] }{\left[ k_{i}^{-}q^{+}-\frac{(k_{i}-k_{2})_{\bot }^{2}+m^{2}}{2x}-\frac{k_{2_{\bot }}^{2}+m^{2}}{2\left( k_{i}^{+}-xq^{+}\right) }q^{+}\right] }  \frac{1}{\left[ k_{f}^{-}q^{+}-\frac{k_{2_{\bot }}^{2}+m^{2}}{2\left( k_{i}^{+}-xq^{+}\right) }q^{+}-\frac{(k_{f}-k_{4})_{\bot }^{2}+m^{2} }{2y}-\frac{(k_{4}-k_{2})_{\bot }^{2}+m_{\sigma }^{2}}{2\left( x-y+1\right) }\right] }\nonumber \\
&&\times \frac{1}{\left[ \left( k_{f}^{-}-k_{i}^{-}\right) q^{+}+\frac{(k_{i}-k_{4})_{\bot }^{2}+m^{2}}{2\left( y-1\right) }-\frac{(k_{f}-k_{4})_{\bot }^{2}+m^{2}}{2y}\right] } \frac{1}{\left[ k_{f}^{-}q^{+}-\frac{k_{4_{\bot }}^{2}+m^{2}}{%
2\left( k_{f}^{+}-yq^{+}\right) }q^{+}-\frac{(k_{f}-k_{4})_{\bot }^{2}+m^{2}%
}{2y}\right] }.  \label{Conveniente 3}  
\end{eqnarray}
\end{widetext}

We observe now that in the substitution of Eq. (\ref{Conveniente 3}) into Eq. (\ref{shif2}) the current component $J_{3}^{\bot }$ is proportional to $\left(q^{+}\right)^{2}$. Therefore, in the Drell-Yan reference frame, the currrent component $J_{3}^{\bot }$, also vanishes. 

In this manner we have shown that for the current components $J_{3}^{-}$, $J_{3}^{+}$ and $J_{3}^{\bot }$, only the contribution for the residue corresponding to the composite system survives in the integration over $k^-$ via Cauchy's formula method. However these components do not contribute to the pair production in the limit $q^{+}\rightarrow 0$.

Having achieved this result we want to generalize it for this specific example, counting the terms that bear $q^{+}$, that is, performing a power counting on the factors $q^+$ for a quick analysis of the result in the frame $q^{+}\rightarrow 0$:
\begin{widetext}
\begin{eqnarray}
\text{Momentum integration} \int dk_{2}^{+}dk_{4}^{+} & \Rightarrow & \left(q^{+}\right)^{2}\int dx dy   \nonumber \\
\frac{1}{\left(k_{i}-k_{2}\right)^{+}k_{2}^{+}\left(k_{i}-k_{4}\right)^{+}k_{4}^{+}\left(k_{f}-k_{4}\right) ^{+}\left(k_{4}-k_{2}\right)^{+}} & \Rightarrow & \frac{1}{\left(q^{+}\right)^{4}}   \nonumber \\
\text{Legs of the type}\:\: \frac{1}{\left[a+\frac{b}{cq^{+}}+\frac{d}{eq^{+}}+...\right]^{4}} & \Rightarrow & \left( q^{+}\right)^{4}  \nonumber \\
\text{Numerator of the type}\:\: \left(a+k_{2{\rm on}}\right) \text{ or }\left(a+k_{4{\rm on}}\right) & \Rightarrow & \left(q^{+}\right) ^{0}=1 \nonumber \\
\text{Numerator of the type} \:\:\left(a+k_{j{\rm on}}\right)\: \text{ with }\:j\neq 2\text{ or }4 & \Rightarrow & \frac{1}{q^{+}} \text{.}
\label{analise}
\end{eqnarray}
\end{widetext}
where $a$ represents the momenta $k_{i}$, $k_{f}$, $k_{f}-k_{i}$, etc. and $b$, $d$, etc.are $k_{2\text{on}}$, $k_{4\text{on}}$ or $k_{j\text{on}}$. We note that as we multiply all these factors together it will always remain at least $\left(q^{+}\right)^{1}$, which in the limit for $q^{+}\rightarrow 0$, makes the current in all regions to vanish. What we conclude here is that the introduction of a virtual boson in comparison to the configuration considered in \cite{Sales-Suzuki}, does not alter the current because the factors in the second and third line of Eq. (\ref{analise}) cancel each other. The important factor is the photon vertex, since this increases the power in the numerator and only with correct factors of $\frac{1}{q^{+}}$ we can cancel the factor coming from the change of variables. 

In a general manner we can count the terms in $q^{+}$ for $n$ intermediate bosons and with one external sources, that is, for $n$ bosons and one photon, we have
\begin{widetext}
\begin{eqnarray}
\text{Momentum integration}\:\: \int  \prod_{j=1}^{n+1}dk_{2j}^{+} & \Rightarrow & \left(q^{+}\right)^{n+1}\int \prod_{j=1}^{n+1}dx_{j}   \nonumber \\
\prod_{j=0}^{n} \frac{1}{\left(k_{i}^{+}-k_{2j+2}^{+}\right) k_{2j+2}^{+}}\prod_{j=1}^{n}\frac{1}{\left( k_{2j+2}^{+}-k_{2j}^{+}\right) \left( k_{f}^{+}-k_{2j+2}^{+}\right) } & \Rightarrow & \frac{1}{\left(q^{+}\right)^{2n+2}}   \nonumber \\
\text{Legs of the type}\:\:\frac{1}{\left[a+\frac{b}{cq^{+}}+\frac{d}{eq^{+}}+...\right] } & \Rightarrow & \left(q^{+}\right) ^{2n+2} \nonumber \\
\text{Numerator of the type}\left(a+k_{2{\text{on}}}\right) \text{or}\left(a+k_{4{\text{on}}}\right) & \Rightarrow &\left( q^{+}\right)^{0}=1   \nonumber \\
\text{Numerator of the type }\left(a+k_{j_{\text{on}}}\right)\:\text{with}\:j\neq 2\text{, }4,...,2n+2 & \Rightarrow & \frac{1}{q^{+}} \text{.} \label{nbosons}
\nonumber
\end{eqnarray}
\end{widetext}

Again we perceive that as we multiply these factors it will always remain a term that is proportional to $\left(q^{+}\right) ^{n}$, and this in the limit $q^{+}\rightarrow 0$, makes the current for all regions to vanish away. Therefore, adding more intermediate bosons makes no difference since the powers in $q^{+}$ that are given by the second and third lines of Eq. (\ref{nbosons}) will always cancel each other. 

Now finally, if we have $m$ external sources for $n$ interacting bosons, we obtain
\begin{widetext}
\begin{eqnarray}
\text{Momentum integration}\:\: \int  \prod_{j=1}^{n+1}dk_{2j}^{+} & \Rightarrow & \left(q^{+}\right)^{n+1}\int \prod_{j=1}^{n+1}dx_{j}   \nonumber \\
\frac{1}{\left(k_{i}^{+}-k_{2}^{+}\right) \cdots \left(k_{2n+2}^{+}-k^{+}_{2n}\right)\left(k_{f}^{+}-k_{2n+2}^{+}\right) \left( k_{f}^{+}-k_{2j+2}^{+}\right) } & \Rightarrow & \frac{1}{\left(q^{+}\right)^{2n+m+2}}   \nonumber \\
\text{Legs of the type}\:\:\frac{1}{\left[a+\frac{b}{cq^{+}}+\frac{d}{eq^{+}}+...\right] } & \Rightarrow & \left(q^{+}\right) ^{2n+m+2} \nonumber \\
\text{Numerator of the type}\left(a+k_{2{\text{on}}}\right) \text{or}\left(a+k_{4{\text{on}}}\right) & \Rightarrow &\left( q^{+}\right)^{0}=1   \nonumber \\
\text{Numerator of the type }\left(a+k_{j_{\text{on}}}\right)\:\text{with}\:j\neq 2\text{, }4,...,2n+2 & \Rightarrow & \frac{1}{(q^{+})^m} \text{.} \label{nbmp} 
\end{eqnarray}
\end{widetext}

In this manner it is only possible to observe the contributions of antiparticles when we put more energy in the system of two interacting bosons. We can check in the case shown previously: In second order of coupling constant for a virtual boson it results in no observation of antiparticle contributions for $q^+ \rightarrow 0$ in a background field. However in the expression Eq. (\ref{nbmp}) we have a case of two external sources ($m=2$) and one interacting intermediate boson ($n=1$) in which we obtain a cancelation od the factors $\frac{(q^+)^{n+1}}{(q^+)^m}=1$. As a consequence, in this case we will have a nonvanishing contribution from the diagrams of antiparticles. Therefore, as we increase the number of photons (more energy input to the system) on the $n$ bosons, we will encounter nonvanishing contributions from pair production diagrams in the limit $q^+ \rightarrow 0$. 

We have plotted in the following figures the pair production contribution in the components of the electromagnetic current in the light front for the propagation of two scalar bosons with one scalar boson exchange. Over this system there is a background field up to to photons (labeled external sources). We can see that for one external photon there is no pair contribution for any of the components of the current. With two photons, there is a pair contribution in the $J^-$ component of the current. The other components, $(J^+, J^\perp)$ have no contribution in any order up to order 2 in the background field. 
\begin{figure}[h]
\begin{center}
\includegraphics[scale=.8]{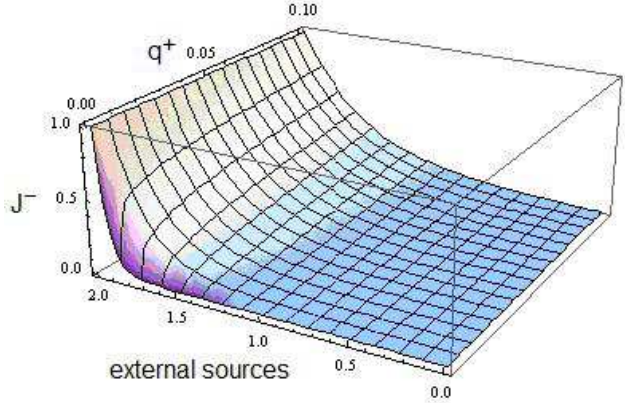}
\caption{$J^-$ Current component in the light-front.}
\label{fig:1} 
\end{center}      
\end{figure}
\begin{figure}[h]
\begin{center}
\includegraphics[scale=.8]{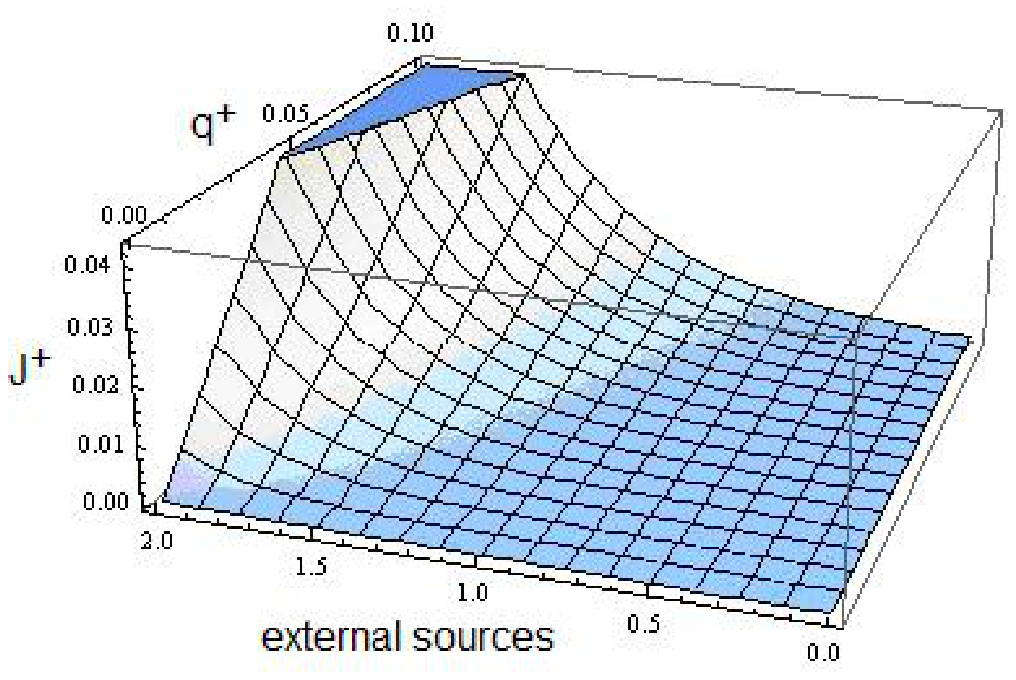}
\caption{$J^+$ Current component in the light-front.}
\label{fig:2} 
\end{center}      
\end{figure}
\begin{figure}[h]
\begin{center}
\includegraphics[scale=.8]{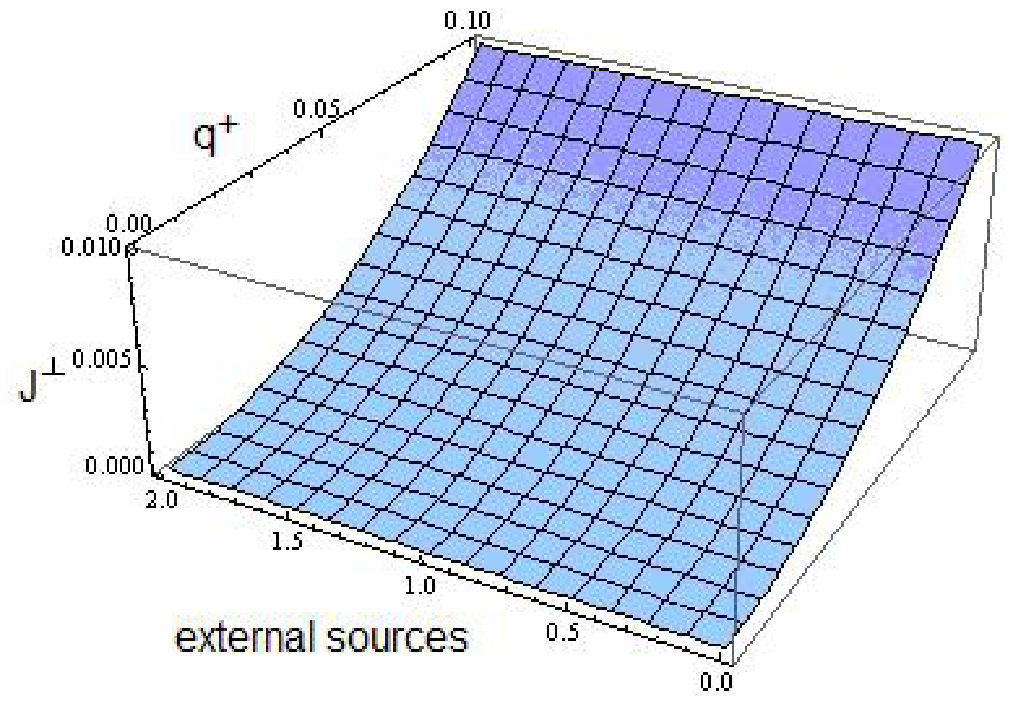}
\caption{$J^\perp$ Current component in the light-front.}
\label{fig:3} 
\end{center}      
\end{figure}

\section{Conclusion}

We have demonstrated that the propagator of two bosons in a background field, has a non-vanishing contribution coming from the pair creation by the photon. In particular, in an example of bound state with constant vertex, we demonstrated that the $J^{-}$ current component in the Breit's reference frame ($q^{+}=0$) has a non-zero contribution from the process of pair
creation by the photon. This conclusion is reached as long as we first have $q^{+}$ different from zero, integrating in $k^{-}$ and then taking the limit $q^{+}\rightarrow 0$. The integration in $k^{-}$ and the limit $q^{+}\rightarrow 0$ does not commute in general.

In the process of these calculations it has been pointed out that the emergence of a non-vanishing contribution from pair production by the interacting photon is naturally achieved by extending the region of allowed quanta solutions in the light-front, that is, extending the Fock space of positive quanta to include relevant solutions from the Fock space of
negative quanta. It also means that the myth of light front trivial vacuum must be forever abandoned.

It has been demonstrated that the inclusion of the pair production term in the light front formalism is of capital importance for the validity of rotational symmetry for the electromagnetic current of a bound state of two bosons in the model of a constant vertex \cite{9}. In the case of components $J^{+}$ and $\vec{J}_{\perp }$ we concluded that the pair
creation term does not contribute in the limit $q^{+}\rightarrow 0$. For the $J^{-}$ component, however, we have shown that we must take into account the pair production so that rotational symmetry be satisfied in the limit $q^{+}\rightarrow 0$. This result has already been known for a while in the light-front \textsl{milieu, }but with our new approach we have shown that
that result which has been reached before via "ad hoc" mathematical techniques, can be achieved on the basis of physical grounds.

We also show thus that the method of ``dislocating the integration pole'' is nothing more than a particular case of our approach, so that such an ``ad hoc'' prescription can be better understood as we deal with the whole Fock space. With this we can also prepare our way to deal with cases involving interactions.

In this work we performed the calculations for corrections to the propagator in a background field up to second order in the coupling. We obtained more diagrams than it was considered in a recent article \cite{Marinho}, just those in which antiparticles appear. The ``Z-graph'' appears naturally in our approach. Yet in the Breit's reference frame these diagrams do not contribute to the current in order $g^{2}$.

For orders in $g^{n}$, perhaps it may be possible to devise a recipe on how to introduce correctly the orders in $q^{+}$ so that the results in some regions survive, as in \cite{Sales-Suzuki} in the Breit's frame.

\begin{acknowledgments}
ATS wishes to thank Southern Adventist University for the kind hospitality, JHS thanks thanks Centro de Armazenamento de Dados e Computa\c{c}\~{a}o Avan\c{c}ada da UESC, and LAS thanks Capes for full financial support.
\end{acknowledgments}

\appendix

\section{Current for two free bosons}

To describe the electromagnetic current for a system composed of two free
bosons, we study the process in which two bosons of the same mass $m$
propagate forward in time and in a given instant in the light front $\bar{x}%
^{+}$ one of them interacts with an electromagnetic field. In the following
we calculate the components of two non interacting boson current in an
external electromagnetic field, when total momenta before and after the
absorption of photon being $K_{i}^{+}>0$ and $K^{+}>0$ respectively.

The Lagrangean density that involves scalar field and electromagnetic field
in interaction is given by 
\begin{equation}
\pounds =D_{\mu }\phi D^{\mu }\phi ^{\ast }-m^{2}\phi ^{\ast }\phi .
\label{3.0.0}
\end{equation}

The derivative between the scalar field and the electromagnetic field is
contained in the covariant derivative $D_{\mu}\phi $.

In the calculation of the propagator for a particle in a background field we
use the interaction Lagrangean of a scalar field and electromagnetic field.
As we have already mentioned, the interaction between the scalar and
electromagnetic field is contained in the first term of (\ref{3.0.0}), so
that the interaction Lagrangean is 
\begin{equation}
\pounds _{\text{I}}=ieA^{\mu }\left( \phi \partial _{\mu }\phi ^{\ast }-\phi
^{\ast }\partial _{\mu }\phi \right) +e^{2}A^{\mu }A_{\mu }\phi \phi ^{\ast
}.  \label{3.1.0}
\end{equation}

The Lagrangean (\ref{3.1.0}) shows immediately that there are two types of
vertices. The first term corresponds to a vertex containing a photon and two
scalar particles. The second vertex contains two photons and two scalar
particles.

Using the concept of generating functional $Z\left[ J\right] $, or vacuum-vacuum transition amplitude in the presence of a external source $J\left( x\right) $, we write 
\begin{eqnarray}
Z\left[ J\right] &=&\int \mathcal{D}\phi\, {\rm e}^{ i\int d^{4}x\left[\pounds \left( \phi \right) +J\left( x\right) +\frac{i\varepsilon }{2}\phi \right] }  \label{3.1.1}  \\
&\varpropto &\left\langle 0,\infty \right. \left\vert 0,-\infty \right\rangle ^{J} 
\nonumber
\end{eqnarray}
where $\pounds =\pounds _{\text{0}}+\pounds _{\text{I}}$ and 
\[
\pounds _{\text{0}}=\partial _{\mu }\phi \partial ^{\mu }\phi ^{\ast
}-m^{2}\phi ^{\ast }\phi 
\]

The Green functions are the expectation values of the time ordered product
of field operators in vacuum and can be written in terms of functional
derivatives of the generating functional $Z_{0}\left[ J\right] $. That is: 
\begin{equation}
G\left( x_{1},...,x_{n}\right) =\left\langle 0\right\vert T\left( \phi
\left( x_{1}\right) ...\phi \left( x_{n}\right) \right) \left\vert
0\right\rangle  \label{3.2.1}
\end{equation}
which are the $n$-point Green functions of the theory, where 
\begin{equation}
\left\langle 0\right\vert T\left( \phi \left( x_{1}\right) ...\phi \left(
x_{n}\right) \right) \left\vert 0\right\rangle =\frac{1}{i^{n}}\left. \frac{%
\delta ^{n}Z_{0}\left[ J\right] }{\delta J\left( x_{1}\right) ...\delta
J\left( x_{n}\right) }\right\vert _{J=0}.  \label{3.2.3}
\end{equation}

Green functions for field theories are extremely important because they are
intimately related to the matrix elements of the scattering matrix $S$ from
which we can calculate quantities measured directly in the experiments such
as scattering processes where the cross section for a given reaction is
measured, decay of a particle in two or more where we can measure the mean
life of particles involved, etc..

The propagator is associated to the Green function equation as: 
\begin{equation}
G(t-t^{\prime })=-iS(t-t^{\prime }).  \label{gpro}
\end{equation}

The Green function or the propagator describes completely the evoltuion for
the quantum system. In this present case we are using the propagator for
``future times''. We could also have defined the propagator ``backwards'' in
time.

The propagation of a free particle with spin zero in four dimensional
space-time is represented by the covariant Feynman propagator 
\begin{equation}
S(x^{\mu })=\int \frac{d^{4}k}{\left( 2\pi \right) ^{4}}\frac{ie^{-ik^{\mu
}x_{\mu }}}{k^{2}-m^{2}+i\varepsilon },  \label{cov}
\end{equation}
where the coordinate $x^{0}$ represents the time and $k^{0}$ the energy. We
are going to calculate this propagation in the light front, that is, for
times $x^{+}$.

We make the projection of the propagator for a boson in time associated to
the null plane rewriting the coordinates in terms of time coordinate $x^{+}$
and the position coordinates $(x^{-}$ and $\vec{x}_{\perp })$. With these,
the momenta are given by $k^{-}$, $k^{+}$ and $\vec{k}_{\perp }$, and
therefore we have 
\begin{equation}
S(x^{+})=\frac{1}{2}\int \frac{dk_{1}^{-}}{\left( 2\pi \right) }\frac{ie^{ 
\frac{-i}{2}k_{1}^{-}x^{+}}}{k_{1}^{+}\left( k_{1}^{-}-\frac{k_{1\perp
}^{2}+m^{2}}{k_{1}^{+}}+\frac{i\varepsilon }{k_{1}^{+}}\right) }.
\label{lf1}
\end{equation}

The Jacobian of the transformation $k^{0}$, $\vec{k}\rightarrow k^{-},k^{+},%
\vec{k}_{\perp }$ is equal to $\frac{1}{2}$ and $k^{+}$, $k_{\bot }$ are
momentum operators.

Evaluating the Fourier transform, we obtain 
\begin{equation}
\widetilde{S}(k^{-})=\int dx^{+}e^{\frac{i}{2}k^{-}x^{+}}S(x^{+}),
\label{transf}
\end{equation}%
where we have used 
\begin{equation}
\delta (\frac{k^{-}-k_{1}^{-}}{2})=\frac{1}{2\pi }\int dx^{+}e^{\frac{i}{2}%
\left( k^{-}-k_{1}^{-}\right) x^{+}},  \label{delta}
\end{equation}%
and the property of Dirac's delta \textquotedblleft
function\textquotedblright\ 
\begin{equation}
\delta \left( ax\right) =\frac{1}{a}\delta \left( x\right) ,  \label{prop}
\end{equation}%
and we get 
\begin{equation}
\widetilde{S}(k^{-})=\frac{i}{k^{+}\left( k^{-}-\frac{k_{\perp }^{2}+m^{2}}{%
k^{+}}+\frac{i\varepsilon }{k^{+}}\right) },  \label{prop1}
\end{equation}%
which describes the propagation of a particle forward to the future and of
an antiparticle backwards to the past. This can be oberved by the
denominator which hints us that for $x^{+}>0$ and $k^{+}>0$ we have the
particle propagating forward in time of the null plane. On the other hand,
for $x^{+}<0$ and $k^{+}<0$ we have an antiparticle propagating backwards in
time.


\begin{thebibliography}{99}

\bibitem{dirac} P.A.M. Dirac, Rev. Mod.Phys. \textbf{21} (1949) 392.

\bibitem{9} J.P.B.C.de Melo, J. H. O. Sales, T. Frederico e P. U. Sauer,
Nucl. Phys. \textbf{A} \textbf{631} (1998) 574c.

\bibitem{37} M. Sawicki, Phy. Rev. \textbf{D44} (1991)433; Phys. Rev. 
\textbf{D46} (1992) 474.

\bibitem{33} S.-J. Chang e T.-M. Yan, Phys. Rev. \textbf{D7} (1973) 1147;
T.-M. Yan, \textit{ibid}. \textbf{D7} (1973) 1780.

\bibitem{34} N.E.Ligterink e B.L.G. Bakker, Phys. Rev. \textbf{D52} (1995)
5917; Phys. Rev. \textbf{D52} (1995) 5954.

\bibitem{35} N.C.J. Schoonderwoerd e B.L.G. Bakker, Phys. Rev. \textbf{D57}
(1998) 4965; Phys. Rev.\textbf{D58} (1998) 025013.

\bibitem{51} J.P.B.C.de Melo, T. Frederico, H.W.L. Naus e P.U.Sauer,
Nucl.Phys. \textbf{A660} (1999) 219.

\bibitem{46a} J.P.B.C.de Melo e T.Frederico, Phys. Rev. \textbf{C55 }%
(1997)2043.

\bibitem{41} S.J.Brodsky, C.R.Ji e M.Sawicki, Phys. Rev. \textbf{D32}
(1985)1530.

\bibitem{38} T. Frederico e G.A.Miller, Phy. Rev. \textbf{D45} (1992)4207;
Phys. Rev. \textbf{D50} (1994) 210.

\bibitem{40} W.Jaus, Phys. Rev. \textbf{D41} (1990)3394.

\bibitem{39} C.M.Shakin e W.-D. Sun, Phys. Rev. \textbf{C51 }(1995)2171.



\bibitem{Sales-Suzuki} J.H.O. Sales, A.T. Suzuki, Int.J.Theor.Phys.\textbf{48%
}, 2340 (2009).

\bibitem{Sales-Suzuki2} J.H.O. Sales and A.T. Suzuki, Int.J.Theor.Phys. 
\textbf{48}: 3173, (2009).

\bibitem{Marinho} J. A. O. Marinho, T. Frederico, P. U. Sauer, Phys. Rev. D 
\textbf{76}, 096001 (2007).

\bibitem{JHS} J.H.O.Sales, T. Frederico, B.V. Carlson and P.U. Sauer, Phys.
Rev. C \textbf{61} (2000) 044003.

\bibitem{jhs2001} J.H.O.Sales, T. Frederico, B. V. Carlson, P.U. Sauer,
Phys. Rev. \textbf{C} \textbf{63}, (2001) 064003.

\bibitem{chine} J. H. O. Sales and Alfredo Takashi Suzuki, Communications in
Theoretical Physics \textbf{55} (2011) 1029.Amsterdam 1973.

\bibitem{bethe} E.E. Salpeter e H.A. Bethe, Phys.Rev. \textbf{84} (1951)
1232.

\bibitem{2} J.M.Namyslowski, Progress in Particle and Nuclear Physics 
\textbf{14 }(1985) 49.

\bibitem{36} S.J.Brodsky, H.-C.Pauli and S.Pinski, Phys.Rep.\textbf{301}
(1998) 299.

\bibitem{Drell} S.D. Drell and Tung-Mow Yan, Phys. Rev. Lett. \textbf{24}
(1970) 181-185.
\end{thebibliography}
\end{document}